\documentclass[aps,prb,twocolumn,showpacs,floatfix]{revtex4}
\usepackage{verbatim} % comment
\usepackage{amsmath,amssymb,graphicx}
\usepackage{epsfig}
\usepackage{graphicx}
\usepackage{enumerate}
\usepackage[toc,page,header]{appendix}

\newcommand{\beq}{\begin{equation}}
\newcommand{\eeq}{\end{equation}}

\newcommand{\bea}{\begin{eqnarray}}
\newcommand{\eea}{\end{eqnarray}}

\begin{document}

\title{Magnetotransport in Aharonov Bohm interferometers: Exact numerical simulations}
\author{Salil Bedkihal and Dvira Segal}
\affiliation{Chemical Physics Theory Group, Department of Chemistry, University of Toronto, 80 Saint George St. Toronto, Ontario, Canada M5S 3H6}


%03.65.Yz, %Decoherence; open sysetems; quantum statistical methods
%73.63.Kv,%  Quantum dots
%73.23.-b,
% Electronic transport in mesoscopic systems
%73.40.-c,
%Electronic transport in interface structures
%73.63.Nm,
%Quantum wires
%73.63.-b,
%Electronic transport in nanoscale materials and structures
%85.65.+h,
%Molecular electronic devices
% 05.70.Ln  Nonequilibrium and irreversible thermodynamics
% 05.30.-d 	Quantum statistical mechanics (
%05.60.-k 	Transport processes

\begin{abstract}
The linear conductance of a two-terminal Aharonov-Bohm interferometer is an even function of the applied
magnetic flux, as dictated by the Onsager-Casimir symmetry. Away from linear response this symmetry may be
broken when many-body interactions are in effect.
%electron-electron interactions are included or inelastic scattering effects are included.
Using a numerically-exact simulation tool, we study the dynamics and the steady-state behavior of the
out-of-equilibrium double-dot Aharonov Bohm interferometer, while considering different types of
interactions: Model I includes a closed interferometer with an inter-dot electron-electron repulsion energy.
In model II  the interferometer is interacting with a dissipative environment, possibly driven away from
equilibrium.
In both cases we show that depending on  the (horizontal, vertical) mirror symmetries of the setup,
nonlinear transport coefficients obey certain magnetosymmetries. We compare numerically exact simulations to
phenomenological approaches:
The behavior of model I is compared to self-consistent mean-field calculations. Model II, allowing heat
dissipation to a thermal bath, is mimicked by an Aharonov Bohm junction with a voltage probe. In both cases
we find that phenomenological treatments capture the relevant transport symmetries, yet significant deviations
in magnitude may show up.
%
%with numerical simulation that while the Onsager-Casimir symmetry is invalidated,
%nonlinear transport coefficients obey certain symmetries: (i)
%When the setup is centrosymmetric, even (odd) conductance terms obey an odd (even) symmetry
% with the threading magnetic flux. (ii) When spatial asymmetry is
%introduced, the above symmetry is respected
%only when the interferometer's geometry is furthermore mirror inverted.
%We also investigate the role of nonequilibrium environments on transport symmetries,
%and demonstrate that such effects do not alter our main conclusions.
\end{abstract}

\date{\today}

\maketitle

%======================================================================
\section{Introduction}

% intro
Microreversibility dictates linear response properties such as the Onsager-Casimir symmetry
relations. Particularly, in a two-terminal conductor the linear conductance should be an even function of
the magnetic field $B$ \cite{OnsagerC}. In Aharonov-Bohm (AB) interferometers, this symmetry is displayed by
the ``phase rigidity" of the conductance oscillations with $B$ \cite{Imry,Yacoby}.
Microreversibility is broken beyond linear response, thus %it is expected that
magnetoassymetries should develop at finite bias, as  demonstrated in several experiments
 \cite{Linke3,Linke1,Linke2,breakE1,breakE2,breakE3,breakE5,Hernandez11}.
What is then remarkable is not the failure of the Onsager symmetry away from equilibrium, rather the
development of more general symmetries between nonlinear transport coefficients and high order cumulants
\cite{Saito08,Saito09,Saito10,Saito11}.
%and the partial restoration of magnetic-field symmetries far-from-equilibrium .
%
%Such general relations may be established beyond linear respone due to the
%existance of other symmetries in the system: particle-hole symmetry or spatial symmetry.
%Validating certain magnetic field symmetries beyond linear response, within  a genuine many-body model, is the
%focus of this paper. We focus on three different models, (a) two terminal closed interferometer with interdot interaction, (b) two terminal
%interferometer coupled capacitively coupled to equilibrium and non-equilibrium Fermionic environment, (c) two terminal %interferometer coupled to
%transitions in the non-equilibrium Fermionic environment.

% calc
Several studies explored magnetotransport in AB interferometers beyond the noninteracting limit, coupling
electrons to either internal or external degrees of freedom \cite{break1, break2, break3, break4, Kubo, Meir}. 
For example, the problem has been explored by implementing mean-field arguments within scattering theories,
focusing on effective quantities such as the screening potential developing in the interferometer in response
to an external bias \cite{break1,Saito09}. At this level, one
can show that magnetoasymmetries develop since internal potentials (the result of many-body interactions)
are asymmetric in the magnetic field away from equilibrium \cite{break1,Saito09}.
In the complementary (phenomenological) B\"uttiker's probes approach \cite{Buttiker} elastic
 and inelastic scattering effects are introduced via probes whose parameters reflect the response of the
conductor to the applied magnetic field and the voltage bias \cite{ABs}.
%This technique has been adopted in Ref. \cite{ABs} for the study of magnetotransport in dissipative 
%AB interferometers.
Beyond phenomenological treatments,
magnetotransport characteristics were investigated using microscopic models in the coulomb blockade limit
\cite{Meir,break4} and in the Kondo regime \cite{break4}.

\begin{figure}[htbp]
\hspace{0mm}
{\hbox{\epsfxsize=55mm \epsffile{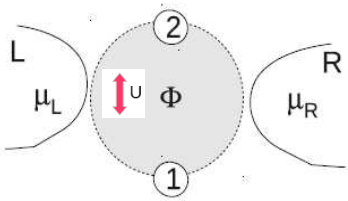}}}
\caption{Scheme of model I, a two-terminal double-dot Aharonov-Bohm interferometer with 
spinless electrons in two quantum dots, $1$ and $2$, with an inter-dot repulsion of strength $U$.} 
\label{schemeI}
\end{figure}

\begin{figure}[t]
\hspace{-16mm}
{\hbox{\epsfxsize=130mm \epsffile{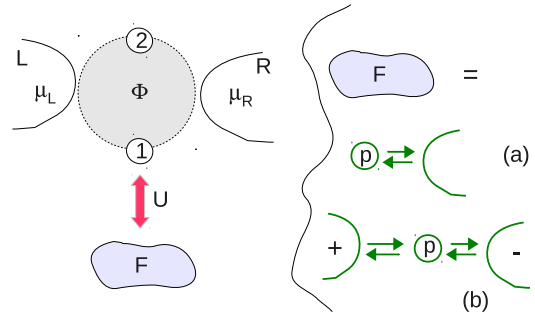}}}
\vspace{-55mm} \hspace{-12mm}
\caption{Scheme of model II, a two-terminal double-dot Aharonov-Bohm interferometer
coupled to a fermionic environment. This environment consists a quantum dot
(labeled $p$) itself hybridized with either (a) an equilibrium sea
of noninteracting electrons, or (b) two metals ($\pm$), possibly biased away from equilibrium.
In both cases dot '1' of the interferometer 
is coupled capacitively (strength $U$) to dot $p$ in the fermionic environment.} 
\label{schemeII}
\end{figure}

In this work we study characteristics of nonlinear transport in AB interferometers by means of an exact
numerical technique \cite{IF}. 
%This tool allows us to explore the time evolution of the charge current, and
%other observables, in a class of interacting systems \cite{IF}.
%
Our setup includes an AB interferometer with two quantum dots, one at each arm, and
we introduce different types of many-body effects within the system: 
Model I includes an inter-dot Coulombic repulsion term, see Fig. \ref{schemeI}. In model II a
secondary fermionic environment interacts capacitively with one of the quantum dots. This environment can 
serve as a ``charge sensor" or a ``quantum point contact", see Fig. \ref{schemeII}.
We simulate the dynamics and the steady-state properties of these (many-body out-of-equilibrium) setups by
adapting an iterative influence functional path integral technique (INFPI), developed in Ref.
 \cite{IF} to treat the dynamics of the single impurity Anderson dot model.

%
%made of a
%quantum dot labeled by $p$,
% possibly representing an adatom or a structured density of states for the metals.
%The dot $p$ is coupled to one or more reservoirs ($\alpha=\pm$).
%When this environment is biased out-of-equilibrium, it
%serves
%as a  ``charge sensor", the current flowing between the $\pm$ metals is
%sensitive to the occupation of electrons in dot 1. Henceforth we refer to the QD labeled by $p$
%and its $\pm$ reservoirs as a ``Fermionic Environment" (FE).
%This setup provides the flexibility of preparing the FE in equilibrium,
%or away from thermal equilibrium.
% compare to MF!!
%We consider two models for AB interferometers, see Figs. \ref{modelI} and \ref{modelII}.
%In the first case inter-dot electron-electron i

Our work includes the following contributions: (i) We study symmetries of magnetotransport
far-from-equilibrium in the transient domain and in the steady-state limit including {\it genuine} many-body
interactions, rather than using phenomenological (screening, probe) models \cite{break1,Saito09}. (ii)
Magnetotransport characteristics were explored in the literature within different models, e.g., considering
an interferometer made with one or two quantum dots, with or without thermal dissipation effects. Here, we
study transport symmetries in different models using the same computational tool, allowing for a
direct comparison. (iii) We compare exact simulations to phenomenological treatments, for clarifying the
validity and accuracy of approximate techniques in magnetotransport calculations.
Particularly, in model II our simulations reveal functionalities beyond the mean field level:
diode (dc-rectification) effect at zero flux under spatial asymmetry and a finite coulomb drag current, driven by the fermionic environment.
%
%(iii) We demonstrate the tunability of magnetoasymmetries with  AB parameters:
%ot energies, temperature and repulsion effects. %  check!
%(iii) We demonstrate that while in non-centroassymetric systems no general symmetries are obeyed,
%odd conductance terms show only a {\it small violation} of the Onsager symmetry, supporting experimental
%results \cite{breakE2}.
%(v) Our work in support of the phenomenological modeling of Ref. \cite{ABs},

The paper is organized as follows. In Sec. II we introduce two models of an interacting double-dot
interferometer, and the principles of the numerical techniques adopted in this work. Results are
presented in Sec. III.  Sec. IV concludes. For simplicity, we set $e$=1, $\hbar=1$, and $k_B=1$.

%===========================

%===========================
%======================================================

\section{Models and Techniques}

\subsection{Double-dot AB interferometer}
We begin with the noninteracting Hamiltonian, common to the two models.
It includes a two-terminal ($\nu=L,R$) AB interferometer with two dots, $n=1,2$, one at each arm.
For simplicity, we ignore the spin degree of freedom (absorbing Zeeman splitting into
the definition of the energies), and take into account only one electronic level in each dot.
The Hamiltonian includes the following terms
\bea
&&H_{AB}= \sum_{n=1,2}\epsilon_na_n^\dagger a_n+
\sum_{l\in L}\epsilon_la_{l}^{\dagger}a_{l} + \sum_{r\in R}\epsilon_ra_{r}^{\dagger}a_{r}
\nonumber\\
&&+ \sum_{n=1,2}\sum_{l\in L} v_{n,l} a_{n}^{\dagger}a_{l}
e^{i\phi_{n}^{L}}
+\sum_{n=1,2}\sum_{r\in R}v_{n,r}a_{r}^{\dagger}a_{n}
e^{i\phi_{n}^{R}}+h.c.
\label{eq:HAB}
\nonumber\\
\eea
Here $a_{k}^{\dagger}$ ($a_{k}$) are fermionic creation (annihilation) operators of electrons with
momentum $k$ and energy $\epsilon_k$ in the $k\in \nu$ metal, $a_{n}^{\dagger}$ and $a_{n}$ are the respective
operators for electrons on the dots, $\epsilon_{n}$ denotes the energy of spin-degenerate levels. The
parameter $v_{n,l}$ stands for the coupling strength of dot $n$ to the $l$ state of the $L$ metal. A similar
definition holds for $v_{n,r}$. These coupling terms are absorbed into the definition of the hybridization
energy
\bea
\gamma_{\nu,n}(\epsilon)=2\pi\sum_{j\in\nu}|v_{n,j}|^2\delta(\epsilon-\epsilon_j).
\label{eq:hyb1}
\eea
In our simulations we set $\gamma_{\nu,n}$,  use a constant density of states for the metals,
up to a sharp cutoff $\pm D$,
then construct the real-valued tunneling elements $v_{n,j}$ by using Eq. (\ref{eq:hyb1}).
The AB phase factors  $\phi_{n}^{L}$ and $\phi_{n}^{R}$ are acquired by electrons in a magnetic field
perpendicular to the device plane. These phases are constrained to satisfy
\beq
\phi_{1}^{L}-\phi_{2}^{L}+\phi_{1}^{R}-\phi_{2}^{R}=\phi =2\pi\Phi/\Phi_{0},
\eeq
and we adopt the gauge
\bea
\phi_{1}^{L}-\phi_{2}^{L}=\phi_{1}^{R}-\phi_{2}^{R}=\phi/2.
\eea
Here $\Phi$ is the magnetic flux threading through the AB ring,
$\phi=2\pi\Phi/\Phi_0$ the magnetic phase, and
$\Phi_0=h/e$ the magnetic flux quantum.
We voltage-bias the AB interferometer, $\Delta \mu\equiv\mu_L-\mu_R$, with $\mu_{L,R}$ as the chemical
potential of the metals; we  use the convention of a positive current flowing left-to-right. We bias the
system in a symmetric manner, $\mu_L=-\mu_R$, but this choice does not limit the generality of our
discussion since the dots may be gated away from the so called ``symmetric point" at which
$\mu_L-\epsilon_{n}=\epsilon_{n}-\mu_R$.

The Hamiltonian (\ref{eq:HAB}) does not include interactions and one can readily obtain the exact form of the
(steady-state) charge current flowing between the terminals,
an even function of the magnetic flux \cite{ABs}. Assuming for simplicity
that the two quantum dots are evenly coupled to the $\nu$ terminal, $\gamma_{\nu}\equiv\gamma_{\nu,n}$,  we
find that \cite{ABsG}
\begin{widetext}
\bea
I(\phi)=
\frac{1}{2\pi }\int_{-\infty}^{\infty}
d\epsilon\frac{4\gamma_L\gamma_R \left[(\epsilon-\epsilon_d)^2\cos^2\frac{\phi}{2} +\left(\frac{\Delta \epsilon}{2}\sin\frac{\phi}{2}\right)^2\right]
[f_L(\epsilon)-f_R(\epsilon)]}
{\left[(\epsilon- \epsilon_d)^2-\frac{\Delta\epsilon^2}{4}-\gamma_L\gamma_R\sin^2\frac{\phi}{2}\right]^2+(\gamma_L+\gamma_R)^2(\epsilon-\epsilon_d)^2},
\nonumber\\
\label{eq:curr0}
\eea
\end{widetext}
with $\epsilon_d=(\epsilon_1+\epsilon_2)/2$ and $\Delta\epsilon=\epsilon_1-\epsilon_2$. The current is an
even function in the magnetic flux at finite bias, irrespective of spatial asymmetries 
($\gamma_L\neq \gamma_R$ and $\Delta \epsilon\neq0$).
Using the probe technique, a phenomenological tool for implementing scattering effects, we
had recently proved that this ``phase locking"
 behavior is preserved under elastic dephasing \cite{ABs}. In contrast,
inelastic scattering processes, taken into account with a voltage probe or a voltage-temperature 
probe, break the even symmetry in $\phi$ in the nonlinear transport regime. We now augment
the Hamiltonian (\ref{eq:HAB}) with genuine many-body interactions:
 In model I we add an inter-dot repulsion interaction between electrons, 
overall conserving energy and charge in the interferometer. 
In model II energy exchange with an additional environment is allowed. 
% yet in both cases we confirm 
%that if the vertical mirror symmetry is maintained, the conductances obey certain symmetry relations [Eq. (\ref{eq:Mfs}) below].

%========================
\subsubsection{Model I: Inter-dot Coulomb repulsion}

We complement the Hamiltonian (\ref{eq:HAB}) with a Coulomb repulsion term,
nonzero when both quantum dots are occupied. The resulting  inter-dot coulomb (C) model reads
\bea
H_C= H_{AB} +
%+V_C
%\nonumber\\
%V_C&=&
Un_1 n_2.
\label{eq:modelI}
\eea
For a schematic representation see Fig. \ref{schemeI}. Here $n_1=a_1^{\dagger}a_1$ and
$n_2=a_2^{\dagger}a_2$  are the number operators for the dots. The behavior of the current and the
occupation of the dots in this ``interacting two-level quantum dot model" were investigated in different
works: The case without the threading magnetic field was studied e.g. in Refs. 
\cite{Sindel,Ora,Wacker,IF2d}.
The role of an external magnetic field was examined in different limits, particularly in the
coulomb blockade regime \cite{Gefen}. Recent studies further investigated transient effects, either
analytically, disregarding interactions \cite{Ora-time}, or numerically, considering relatively weak
interactions \cite{Salil11}.
%
%Here we do not repeat these extensive studies, rather only focus on the behavior of $\mathcal R(\phi)$
%and$\mathcal D(\phi)$ when horizontal (left-right) or axial (top-down) mirror symmetries are broken.

In our simulations below we consider three geometries for  Model I, see Fig. \ref{FigS}:
(i)  A setup with a mirror symmetry with respect to
the horizontal axis,
%Here the dots are taken as energy-degenerate, $\epsilon_1=\epsilon_{2}$, and
%we use 
%$\gamma_{\nu}=\gamma_{\nu,1}=\gamma_{\nu,2}$. However, a left-right asymmetry in the form $\gamma_{L,n}\neq
%\gamma_{R,n}$ is included.
%% see Fig. \ref{ModelIINFPI1}.
 (ii) the case with a mirror symmetry along the vertical axis, and
%, $\gamma_{L,n}=\gamma_{R,n}$, but $\epsilon_1\neq\epsilon_2$, and possibly
%$\gamma_{\nu,1}\neq\gamma_{\nu,2}$.
%%, see Fig. \ref{ModelIINFPI3}. 
(iii) the model missing (horizontal and
vertical) symmetries. % see Fig. \ref{ModelIINFPI2}.

%--------------------------------------------------
\subsubsection{Model II: Coupling to a Fermionic environment}

Dissipation effects can be included by capacitively coupling the interferometer to a fermionic environment
(FE), set in equilibrium or out of equilibrium. For simplicity, we do not consider electron-electron
interactions within the interferometer or within the FE. This dissipative (D) Hamiltonian includes the
interferometer  [Eq. (\ref{eq:HAB})], an additional FE, and the interaction energy between the units,
\bea
H_{D}= H_{AB} + H_{F} + V_{D}.
\label{eq:modelII}
\eea
The FE is realized here by a tunneling junction
\bea
H_{F}=\epsilon_{p}c_p^{\dagger}c_p
+ \sum_{s\in \alpha}\epsilon_sc_{s}^{\dagger}c_{s} +
% \sum_{r\in R}c_{r}^{\dagger}c_{r}+
\sum_{s\in \alpha} g_{s}c_{s}^{\dagger}c_p+h.c.
\label{eq:HF}
\eea
%a
It includes a quantum dot of energy $\epsilon_p$
coupled to two reservoirs ($\alpha=\pm$).
The FE may be set at equilibrium when $\mu_+=\mu_-$ (with
the Fermi energy set at zero), or
biased away from  equilibrium,  $\Delta\mu_{F}\equiv \mu_+-\mu_-\neq0$.
We distinguish between the AB interferometer and the FE by adopting the operators $c^{\dagger}$ and $c$ to
denote creation and annihilation operators of electrons in the FE. We define the dot-reservoir hybridization
energies in the FE by
\bea
\gamma_{\alpha}(\epsilon)=2\pi\sum_{s\in\alpha}|g_{s}|^2\delta(\epsilon-\epsilon_s).
\eea
Electrons in the AB interferometer and the FE are interacting (strength $U$)
according to the form
\bea
V_D=Un_{p}n_1.
\label{eq:HIIint}
\eea
Here $n_p=c_p^{\dagger}c_p$, $n_1=a_1^{\dagger}a_1$ are number operators.
Note that there is no leakage of electrons from the AB junction into the FE. 
However, this additional environment
provides a mechanism for inducing elastic and inelastic scattering events of electrons on dot 1.

Model II has been examined in the literature in the context of charge sensing, and as a 
``which-path" detector, see for example Refs. \cite{KuboQPC,QPC1,QPC2}.
The dephasing in an AB
interferometer with a capacitively coupled charge sensor has been analyzed in Ref. \cite{KuboQPC},
using a second-order perturbation theory in $U$, yet
limited to the linear conductance case. Here, using a numerical tool, we analyze the
system away from equilibrium with the vertical mirror symmetry either preserved or violated, see Fig. 
\ref{FigS}.

%=================================
% observables
\subsection{Observables}

The principal observable of interest in our work is the charge current in the interferometer. 
It is calculated from the current operator, %($e/\hbar=1$)
\bea
\hat I_L&=&-\frac{d N_L}{dt} = %-i\frac{e}{\hbar}[H, N_L] =
-i[H, N_L] =
\nonumber\\
 &=&\sum_{l\in L}\left[-iv_{1,l}e^{i\phi_{1}^{L}}a_1^{\dagger}a_{l}
 +iv_{l,1}e^{-i\phi_{1}^{L}}a_{l}^{\dagger}a_1\right],
\eea
with $N_L=\sum_{l\in L}a_{l}^{\dagger}a_{l}$ as the number operator of the $L$ metal. The current at the $R$
contact, $\hat I_R$, can be defined in a analogous way. We identify the averaged current by $\hat I\equiv
\frac{1}{2}(\hat I_L-\hat I_R)$; we could simulate separately the currents at the $L$ and $R$ terminals, but
we chose to directly compute the expectation value 
\bea
I(t,\phi)={\rm tr}[\rho(0) e^{iHt}\hat Ie^{-iHt}],
\label{eq:Idef}
\eea
from a dot-metal-FE factorized initial state $\rho(0)$.
Here $H$ denotes the total Hamiltonian of interest.
%This measure satisfies certain symmetries in the transient regime, as we show below.
%In the steady state limit, we omit the time variable, and
Formally, the two-terminal current $I(t,\phi)$ can be expanded in powers of the applied voltage
bias $\Delta \mu$,
\bea
I(t,\phi)=\sum_{k=1,2,..}G_k(t,\phi)(\Delta \mu)^k.
%G_1(t,\phi) \Delta\mu + G_2(t,\phi) (\Delta\mu)^2  + G_3(t,\phi)(\Delta\mu)^3 +....
\label{eq:II}
\eea
We refer to $G_{k>1}$ as nonlinear conductance coefficients.
The current can be separated into its odd and even terms in powers of the bias.
Even terms represent the dc-rectification contribution,
\bea
\mathcal R(t,\phi)&\equiv& \frac{1}{2}[I(t,\phi)+\bar I(t,\phi)]
\nonumber\\
&=& G_2(t,\phi) (\Delta\mu)^2  + G_4(t,\phi) (\Delta\mu)^4+ ...
\label{eq:R}
\eea
Here $\bar I$ is the current obtained when interchanging the chemical potentials of the two terminals
(assuming identical temperatures). The complementary odd  terms are  grouped  into
\bea
\mathcal D(t,\phi)&\equiv&  \frac{1}{2}[I(t,\phi)-\bar I(t,\phi)]
\nonumber\\
&=&
G_1(t,\phi) \Delta\mu  + G_3(t,\phi) (\Delta\mu)^3+ ... .
\label{eq:D}
\eea
%
%It is also useful to define a measure for the magnetic field asymmetry of the current,
%%
%\bea
%\Delta I(t,\phi)\equiv [I(t,\phi)-I(t,-\phi)]/2.
%\eea
%%
Below we relax the time variable when addressing 
%denote by $\mathcal D(\phi)$ and $\mathcal R(\phi)$ 
long-time quasi steady-state values.
Other quantities of interest are the occupation of the dots in the interferometer and the behavior of the
coherences, off-diagonal terms of the reduced density matrix, obtained from $\sigma_{n,n'}(t)\equiv {\rm
tr_e}[e^{-iHt} \rho(0)e^{iHt}]_{n,n'}$; the trace is performed over degrees of freedom in the electronic
reservoirs. We could also simulate the dynamics within the FE (Model II),
exploring its operation as a quantum point contact and a magnetic field sensor.

\begin{figure}[htbp]
%\hspace{-2mm}
\vspace{10mm}
\rotatebox{-90}
{\hspace{-14mm}
\hbox{\epsfxsize=90mm  \epsffile{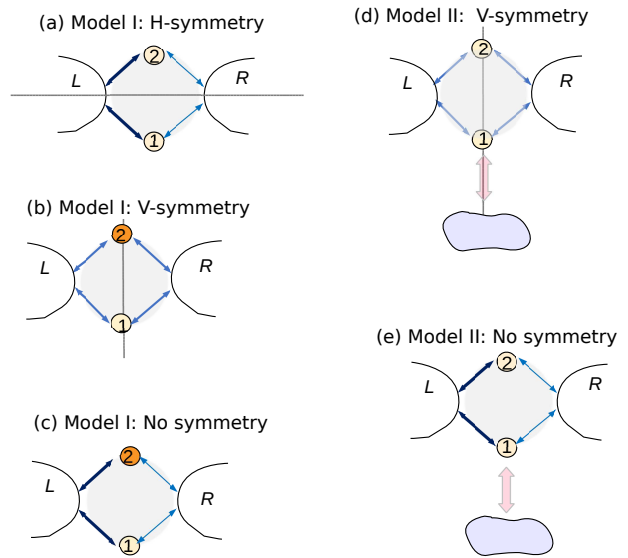}}}
\caption{Schemes of the different setups considered in this work.
(a) Model I with a horizontal (H) mirror symmetry. The bold (light) arrows represent strong (weak) hybridization energies of the quantum dots to the metals. 
Simulations are included in Fig. \ref{ModelIINFPI1}.
(b) Model I with a vertical (V) mirror symmetry. 
The nondegenerate dots are represented by different colors.
%coupled evenly to the metals, 
Fig. \ref{ModelIINFPI3} displays the transients; Fig. \ref{ModelIssI} shows
 steady-state values.
(c) Model I missing the H and V mirror symmetries,  see data in Figs. \ref{ModelIINFPI2} and \ref{ModelIssI}.
(d) Model II with a V mirror symmetry. 
%Note that the model would  maintain its V symmetry  non-degenerate dots.
Results are displayed in Figs. \ref{FE1} and \ref{FE2}.
(e) Model II missing both the H and V mirror symmetries. 
Steady-state data is included in Fig. \ref{FE2}.
} \label{FigS}
\end{figure}

Focusing on the current in the AB unit,  we consider different setups for Model I and II, see Fig. \ref{FigS}.
From symmetry considerations, 
under a mirror symmetry with respect to the horizontal (H) axis, phase locking 
should take place, with $I(\phi)=I(-\phi)$ \cite{Gefen}.
On the other hand, if the system acquires only a mirror symmetry with respect to the vertical (V)
axis, the resulting symmetry relation is $I(\phi)=-\bar I(-\phi)$ \cite{Gefen}, leading to
\bea
\mathcal R(\phi)=-\mathcal R(-\phi),\,\,\,\,
\mathcal D(\phi)=\mathcal D(-\phi).
\label{eq:Mfs}
\eea
Below, our numerical results confirm these relations in the transient domain as well, under
 a symmetrized definition of the current operator, with a symmetric application of bias ($\mu_L=-\mu_R$).
%and with factorized  initial conditions. 
We quantify
the importance of $\mathcal R$ relative to $\mathcal D$, study deviations from the above relations
when the vertical and horizontal symmetries are broken, and test predictions
of approximate techniques against exact simulations.

%===========================

%===========================

\subsection{Numerically Exact Treatment: INFPI}

We simulate the dynamics of electrons in Model I and II using a numerically exact influence function path
integral technique, referred to as INFPI. The principles of this method have been detailed in several recent
publications \cite{IF,IF2d,IFs} therefore we only highlight here the aspects of relevance to the present work.

Eqs. (\ref{eq:modelI}) and (\ref{eq:modelII})
can be each organized into the generic form $H=H_0+H_1$, with $H_0$
comprising the exactly solvable (noninteracting) terms.
Many-body interactions are collected into $H_1$, written in the form
\bea
H_1=U\left[n_1n_a-\frac{1}{2}(n_1+n_a)\right].
\label{eq:H1}
\eea
In model I, $n_a=n_2$; in model II it corresponds to the number operator of the impurity level $p$ within
the FE, $n_a=n_p=c_p^{\dagger}c_p$, see Eq. (\ref{eq:HIIint}). The two-body term $\frac{1}{2}U(n_1+n_a)$ is
absorbed into the definition of $H_0$. This structure allows for the elimination of $H_1$ via the
Hubbard-Stratonovich (HS) transformation and the propagation of quadratic expectation values with an
influence functional path integral technique. We now briefly review the principles of INFPI, to explain why
this method is fitting for the study of magnetotransport in far-from-equilibrium situations. We discuss the
numerical errors associated with INFPI simulations, and point out that these errors do not interfere with
the resolution of transport symmetries.

 The starting point in our approach is the formal expression  (\ref{eq:Idef}).
As an initial condition we use a factorized initial state. For example, in Model II we use
 $\rho(0)=\rho_{F}(0)\otimes \rho_{AB}(0)$ with $\rho_{AB}$ as the state of the interferometer. We further assume that
$\rho_{AB}(0)=\sigma(0)\rho_L \otimes\rho_R$, with $\sigma(0)$ as the reduced density matrix of the isolated
dots in the interferometer. The FE is similarly prepared in a factorized state with
$\rho_{F}(0)=\rho_p(0)\otimes\rho_{+}\otimes\rho_-$. The four reservoirs $\xi=L,R,\pm$ are separately
prepared in grand canonical states at a given chemical potential and temperature,
$\rho_{\xi}=e^{-\beta(H_{\xi}-\mu_{\xi}N_{\xi})}/{\rm
tr}[e^{-\beta(H_{\xi}-\mu_{\xi}N_{\xi})}]$; we prepare all reservoirs at the same temperature
$1/\beta$.

Using this initial state in Eq. (\ref{eq:Idef}), we apply the Trotter decomposition and the HS
transformation, the latter eliminates the many-body term $H_1$ by introducing auxiliary Ising variables \cite{HS}. The
result is a formally exact path integral expression; the integrand is refereed to as the``Influence
Functional" (IF) involving nonlocal dynamical correlations, generally missing an
analytical form. 

The fundamental principle behind INFPI is the observation that at finite temperature and/or nonzero chemical
potential difference bath correlations exponentially decay in time, thus the IF can be truncated beyond a
memory time $\tau_c$ \cite{QUAPI,ISPI}. This allows us to define an auxiliary operator on the time-window $\tau_c$, which can
be time-evolved iteratively from the initial condition to time $t$.
The truncated IF can be evaluated numerically by discretizing the fermionic reservoirs and tracing out
the baths' degrees of freedom using trace identities \cite{IF}.

The INFPI method involves three numerical errors: (i) The finite discretization of the reservoirs,
 each comprising $L_s$ single-electron states. (ii) The time step adopted in the Trotter breakup, $\delta
t$. In our simulations the trotter error grows as $(\delta t U)^2$.  %XXX
(iii) The error associated with the truncation of the IF, to include only a finite memory time $\tau_c$.
The exact limit is reached when $L_s\rightarrow \infty$, $U\delta t \rightarrow 0$ and $\tau_c\rightarrow t$.
Convergence is tested by studying the sensitivity of simulations to the energy discretization of the
reservoirs, the time step, and the memory time
 $\tau_c=N_s\delta t$, with $N_s$ as an integer.

INFPI excellently fits for the 
%over other numerical and analytical techniques,
simulation of magnetotransport in far-from-equilibrium situations: 
First, analytic considerations and numerical
simulations suggest that the memory time characterizing the bath decorrelation process
scales as $\tau_c\sim 1/\Delta \mu$ \cite{IF,ISPI,QUAPI}. Thus, the
method should quickly converge to the exact limit at a large bias. 
Since we are specifically interested here in
beyond-linear-response situations, INFPI is perfectly suited for the problem. Second, 
this is a deterministic time propagation scheme. Thus, it is an advantageous tool for
testing magnetic field symmetries in nonlinear transport:
Even if convergence is incomplete, $I(t,\phi)$ and $I(t,-\phi)$ deviate from the exact limit in the same
(deterministic) form, conserving transport symmetries. 
In contrast, methods that rely on stochastic sampling of diagrams may accumulate
 distinct errors in the evaluation of the current at opposite phases, $I(t,\pm \phi)$, 
thus one may need to approach the exact limit for
validating transport symmetries. 
As we show below, at finite interactions the evolution of the current with time 
strongly depends on the magnetic phase, showing distinct relaxation times.
Thus, it is important to adopt here techniques 
which accumulate identical errors for $\pm \phi$. 
Finally, INFPI is a flexible tool and it can be easily adapted 
 for the study of several related models, as long as the interacting contribution $H_1$ follows Eq.
(\ref{eq:H1}). This allows us to analyze and compare the behavior of different many-body situations, e.g.,
with or without a dissipative bath.

%--------------------------

\subsection{Phenomenological Approaches}

%% More refs for GF
Nonlinear transport characteristics in Model I and II can be explored based on perturbation theory
expansions in $U$, see for example Ref. \cite{KuboQPC}. Alternatively, in mean-field (MF) approaches
many-body effects are embedded in the (noninteracting) scattering formalism. For example, one can write the
scattering matrix as a functional of an electrostatic screening potential which depends on the applied bias
\cite{break1,break3}. Another phenomenological scheme has been explored in Ref. \cite{ABs}, with  
Buttiker's probes used to  implement different processes (elastic, inelastic, dissipative) into the
otherwise coherent dynamics.

In Sec. III we compare results from INFPI to phenomenological methods.
% and show that these
%approximate
%approaches provide the correct symmetries and their violation upon breaking spatial symmetries.
%However,
%they predict absolute values (for example, for
%$\mathcal R$) which substantially deviate from the exact limit. 
%
Model I is analyzed in the
steady-state limit using the self-consistent Hartree approximation as described in Refs. 
\cite{Sindel,Komnik,Zarand}. This scheme accounts for the inter-dot
Coulomb interaction by replacing the bare levels with Hartree energies, e.g., $\epsilon_1\rightarrow
\epsilon_1+U \sigma_{2,2}$. We then use standard  Green's function expressions
 \cite{ABsG}, iterated to reach self-consistency. 

We analyze model II using the phenomenological
voltage probe technique, extended to the nonequilibrium regime, as explained in Ref. \cite{ABs}. 
In this approach we hybridize the quantum dot '1' to a metal terminal (probe), then impose the condition of 
zero net charge current in this connection. 
Electrons can thus dephase and exchange energy in the probe, but (net) charge current only flows
between the $L$ and $R$ terminals.

%=========================
\section{Results}

% mean field <--> exact
% probe <--> exact

In our simulations below we adopt the following parameters:
$\Delta \mu=0.6$, inverse temperature of the electronic reservoirs $\beta=50$, $U=0.1$, 
$\gamma_{\nu,n}=0.05-0.2$, flat bands extending between  $D=\pm1$.
INFPI numerical parameters are $\delta t=0.5-1.2$, $N_s=3-6$, and $L_s=120$.
Convergence was reached for $\tau_c\sim 2$, in agreement with the rough estimate $\tau_c \sim ~1/\Delta \mu$.
We consider different setups, obeying or violating the 
horizontal and vertical mirror symmetries, see Fig. \ref{FigS}.

%===========================
%------------------------------------

\vspace{3mm}
\begin{figure}[htpb]
\hspace{2mm}
{\hbox{\epsfxsize=70mm \epsffile{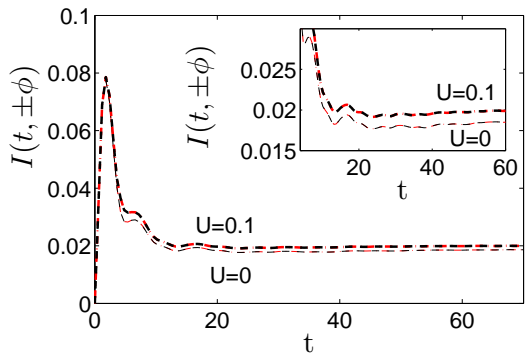}}}
\caption{Model I with a horizontal mirror symmetry, showing phase rigidity. INFPI simulations
using $\phi=\pm\pi/2$, $\gamma_{L,1}=\gamma_{L,2}=0.2$, $\gamma_{R,1}=\gamma_{R,2}=0.05$, $\epsilon_1=\epsilon_2=0.15$,
$\Delta\mu=0.6$,  $U=0$ (bottom two overlapping lines) and $U=0.1$ (top two overlapping lines)
$N_{s}=6$ and $\delta t=0.6$. % set-up C Gefens paper where phase rigidity is obeyed.
The inset zooms on the time evolution after the early transients.
} \label{ModelIINFPI1}
\end{figure}

\begin{figure}[t]
\hspace{2mm}
{\hbox{\epsfxsize=75mm \epsffile{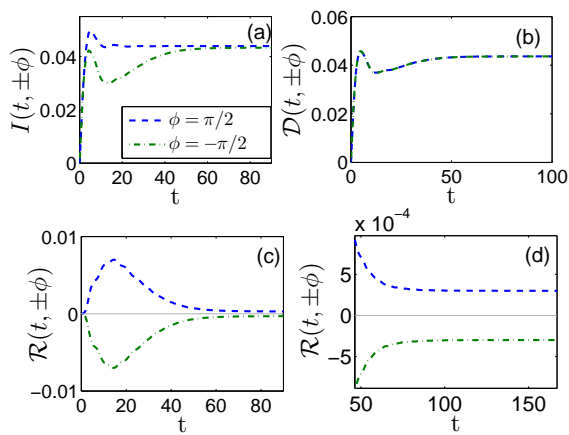}}}
\caption{Model I with a vertical mirror symmetry.
(a) Total current and its breakup into
(b) $\mathcal D$ and (c)-(d) $\mathcal R$ components,
$\phi=\pi/2$,
$\gamma_{\nu,n}=0.1$, $\epsilon_1=0.1$, $\epsilon_2=0.2$,
 $\Delta\mu=0.6$, $U=0.1$, $N_{s}=6$, and $\delta t=0.6$.} %
 \label{ModelIINFPI3}
\end{figure}

\begin{figure}[t]
\hspace{2mm}
{\hbox{\epsfxsize=75mm \epsffile{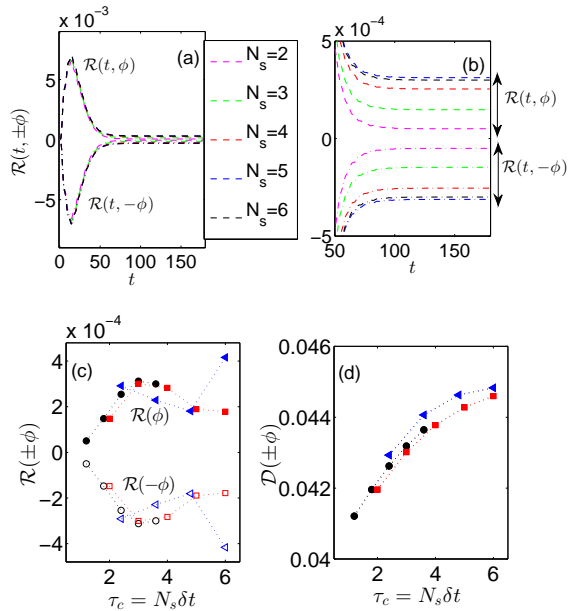}}}
\caption{Example of a convergence analysis for Model I with a vertical mirror symmetry
using the data of Fig. \ref{ModelIINFPI3}.
(a)-(b)  $\mathcal R(t,\pm \pi/2)$ with $\delta t=0.6$ and $N_s=2,3...,6$.
Panel (b) zooms on the long time behavior.
(c)-(d) Long time ($t\sim180$) quasi steady-state values of 
 $\mathcal R(\pm \pi/2)$ and $\mathcal D(\pm \pi/2)$ as a function of $\tau_c=N_s\delta t$
for different time steps,
$\delta t=0.6$ ($\circ$), $\delta t=1$ ($\square$) and $\delta t=1.2$ (triangle).
Filled (empty) symbols stand for the $\phi=\pi/2$ ($\phi=-\pi/2$) magnetic phases,
overlapping in panel (d).
} 
\label{ModelIC}
\end{figure}

\begin{figure}[t]
\hspace{2mm}
{\hbox{\epsfxsize=75mm \epsffile{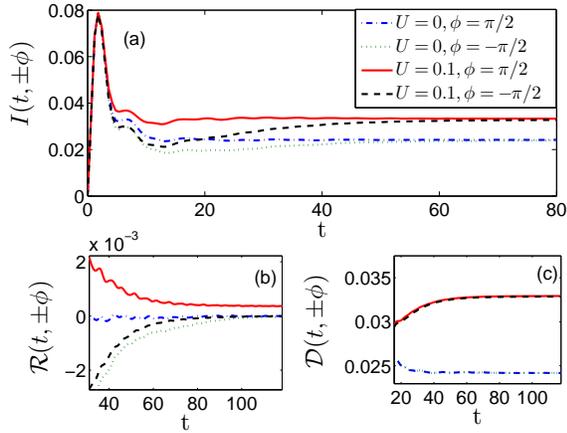}}}
\caption{Model I. (a) Total current and its breakup into its
 (b) $\mathcal R$ and (c) $\mathcal D$ components
for noncentrosymmetric and non-degenerate double dot system,
$\phi=\pi/2$,
$\gamma_{L,1}=\gamma_{L,2}=0.2$, $\gamma_{R,1}=\gamma_{R,2}=0.05$, $\epsilon_1=0.1$, $\epsilon_2=0.2$,
 $\Delta\mu=0.6$, $U=0.1$, $N_{s}=6$ and $\delta t=0.6$.}
 % Broken phase rigidity
 \label{ModelIINFPI2}
\end{figure}

%-----------------------------------------------------

\subsection{Model I}

We analyze nonlinear transport behavior %magnetotransport
in model I using INFPI simulations, then compare our results to the Hartree mean-field approach.
We begin with a setup preserving the horizontal symmetry, $\epsilon_1=\epsilon_2$,
$\gamma_{L,1}=\gamma_{L,2}> \gamma_{R,1}=\gamma_{R,2}$. As pointed out in Ref. \cite{Gefen},
the current in this model should be an even function of the magnetic
field, beyond linear response. 
Fig. \ref{ModelIINFPI1} confirms this ``phase locking" behavior
in the transient regime as well, given the symmetric initial condition and 
the symmetrized definition of the current operator.

In Fig. \ref{ModelIINFPI3} we consider a setup with only a vertical mirror symmetry.
%We use non-degenerate levels but assume identical hybridization constants, $\gamma_{\nu,n}=0.1$.
We separate the current into its odd and even conductance terms by studying the dynamics
with a reversed bias, then calculating $\mathcal R=(I+\bar I)/2$ and $\mathcal D=(I-\bar I)/2$. 
Using INFPI, we find that Eq. (\ref{eq:Mfs}) is obeyed 
even before steady-state is reached.
We exemplify the convergence behavior of this model in Fig. \ref{ModelIC}. In panels (a)-(b)
we display $\mathcal R(t,\pm \pi/2)$ and demonstrate that it obeys the symmetry relation (\ref{eq:Mfs})
before convergence is achieved.
We also examine the steady-state behavior of the system
%and plot $\mathcal R(\pm \pi/2)$ 
%and $\mathcal D(\pm \pi/2)$  
using different values for the simulation time step, see panels (c)-(d).
A large time-step $\delta t=1.2$ does not allow convergence, but with $\delta t=0.6$,
$\mathcal R(\pm \pi/2)$ converges around $\tau_c=3$. \cite{comment2}
Odd conductance terms (panel d) slowly converge,
but maintain transport symmetries (overlapping data for $\mathcal D(\pm \phi/2)$)
under a short memory time.
This observation is not trivial: the dynamics under the phases $\pm\phi$ is quite
different, see for example Fig.  \ref{ModelIINFPI3}(a).

The symmetry relations (\ref{eq:Mfs}) are invalidated
 when the horizontal and vertical mirror symmetries are broken, see Fig.
\ref{ModelIINFPI2}. 
Note that in panel (c), $\mathcal D(\phi)\neq\mathcal D(-\phi)$; deviations are order of $10^{-4}$.
We can use Fig. \ref{ModelIINFPI2} and estimate the magnitude of high order conductances.
For example, from the behavior of $\mathcal D$ (assuming $G_3$ provides the largest contribution after $G_1$) we find
that at $\phi=\pi/2$, $G_1\sim \mathcal D/\Delta \mu =0.06$,
$G_3\sim (\mathcal D(\pi/2)-\mathcal D(-\pi/2))/2\Delta \mu^3\sim5 \times 10^{-4}$ and
$G_2(\pi/2)\sim R(\pi/2)/\Delta\mu^2\sim10^{-3}$.
Thus, at this phase,
 $G_3/G_1=10^{-2}$ and $G_2/G_1=2\times 10^{-2}$.
These conductances are translated to physical
units when multiplied by the factor $\frac{2e^2}{\hbar}$. Note that $G_2$ and $G_3$
are of the same order of magnitude.

\begin{figure}[htpb]
{\hspace{-12mm}
\hbox{\epsfxsize=100mm \epsffile{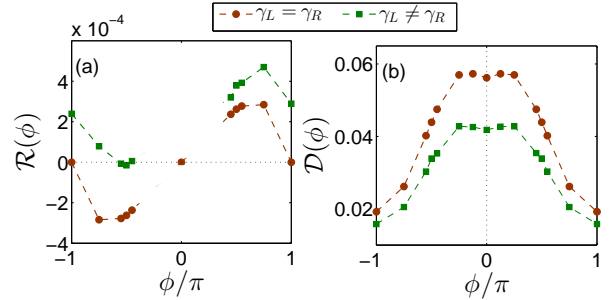}}}
\caption{Steady-state behavior of Model I with INFPI simulations
using nondegenerate dots, $\epsilon_1=0.1$, $\epsilon_2=0.2$.
Setups with a vertical mirror symmetry, $\gamma_{\nu,n}=0.1$ ($\circ$),
and missing mirror symmetries, $\gamma_{L,n}=0.2$ and
$\gamma_{R,n}=0.05$ ($\square$).
Other parameters are
$\Delta\mu=0.6$, $U=0.1$, $\beta=50$,
$N_{s}=6$ and $\delta t=0.6$.}         % Broken phase rigidity but new symmetries
 \label{ModelIssI}
\end{figure}

\begin{figure}[htpb]
{\hbox{\epsfxsize=70mm \epsffile{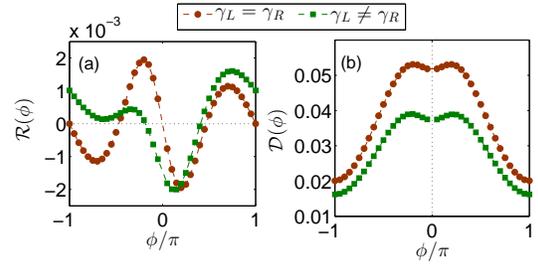}}}
\caption{Steady-state behavior of Model I using a mean-field scheme, and
 the same parameters as in Fig. \ref{ModelIssI}.
Setups with a vertical mirror symmetry, $\gamma_{\nu,n}=0.1$ ($\circ$),
and missing mirror symmetries, $\gamma_{L,n}=0.2$ and
$\gamma_{R,n}=0.05$ ($\square$).}
\label{ModelIssMF}
\end{figure}

In Fig. \ref{ModelIssI} we display the long-time quasi steady-state data 
for $\mathcal R(\phi)$ and $\mathcal D(\phi)$.
Within the present simulation times, we have
 not reached the steady-state limit for $\mathcal R$ using $\phi/\pi<1/2$ 
\cite{comment}.
%auch difficulty did not arise in the calculation of the $\mathcal D$ component.
%
We compare exact simulations to a mean-field
approach as explained in Sec. II.D, see Fig. \ref{ModelIssMF}. 
Both exact and approximate treatments satisfy the relations (\ref{eq:Mfs})
when the vertical mirror symmetry is preserved.
However, the Hartree approach is unreliable as it predicts incorrect magnitudes for $\mathcal R$.
%and the existence of special
%points in which $\mathcal R(\phi)=0$ while $\phi\neq m\pi$; $m$ is an integer, see Fig. \ref{ModelIssMF}(a).
Similarly, in the absence of mirror symmetries, exact and approximate tools demonstrate the violation
of Eq. (\ref{eq:Mfs}), but the Hartree approach overestimates the magnitude of $\mathcal R$.

Prior studies of nonlinear transport in quantum dot systems 
had indicated that the Hartree MF approximation suffers from fundamental artifacts,
e.g.,  it predicts an incorrect hysteresis behavior in the single-impurity Anderson model \cite{Zarand}.
Here we find that the method conserves the correct transport symmetries, but it produces 
incorrect values for the nonlinear terms.
We have also implemented a Hartree-Fock (HF) approach as described in Ref. \cite{Berko}
by further correcting off-diagonal elements in the Green's function with the expectation values
of the coherences $\sigma_{1,2}$. This had reduced the amplitude of
the oscillatory pattern around $\phi/\pi=\pm 0.2$, %(missing altogether in INFPI),
but HF results still overestimate $\mathcal R$ by almost an order of magnitude, for $\phi\sim \pi/2$. 
It is interesting to adopt an equations-of-motion treatment \cite{Haug} 
and explore these deviations maintaining higher-order correlation effects.

We conclude: (i) The Hartree mean-field approach properly describes the development of transport symmetries
when the device acquires vertical or horizontal mirror symmetries; its quantitative predictions are unreliable.
(ii) The AB interferometer can act as a charge diode ($\mathcal R\neq 0$) in a spatially symmetric device
($\gamma_{\nu,n}$ are all identical) if the following conditions are simultaneously met: 
the dots' energies are nondegenerate,
the magnetic flux is nonzero $\phi\neq\pi m $,
 and man-body interactions are in play, see Fig. \ref{ModelIssI}(a). This observation agrees with recent
simulations based on wave-packets propagation \cite{BaowenAB}.

%==================================================================

%-----------------------------------------------------

%------------------------------------

%===============
%MODEL II

\begin{figure}[htbB]
{\hbox{\epsfxsize=80mm \epsffile{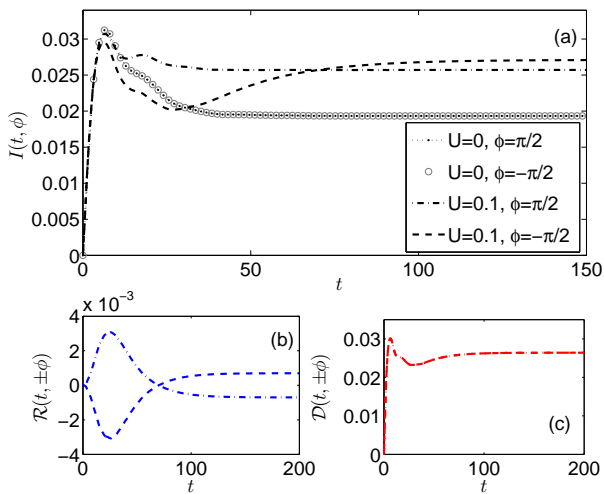}}}
\caption{Model II, an interferometer coupled to an equilibrium FE, with a vertical mirror symmetry.
(a) Charge current in the AB interferometer,
$U=0$ with  $\phi=\pm\pi/2$ (dot and circles, overlapping) and
$U=0.1$ with $\phi=\pi/2$ (dashed-dotted), $\phi=-\pi/2$ (dashed).
(b)-(c)  Odd and even conductance terms
obey the symmetries (\ref{eq:Mfs}).
The quantum dots in the AB interferometer
are set at $\epsilon_{1,2}=0.15$,  $\gamma_{\nu,n}=0.05$.
The FE is set at equilibrium ($\mu_{\pm}=0$) with $\epsilon_p=-0.5$ and
$\gamma_{\pm}=0.2$. The four reservoirs are prepared at the temperature $1/\beta=50$.
Numerical parameters are $\delta t=0.6$,
$N_s=4$ and $L_s=120$.}
\label{FE1}
\end{figure}

\begin{figure}[htbp]
{\hbox{\epsfxsize=70mm \epsffile{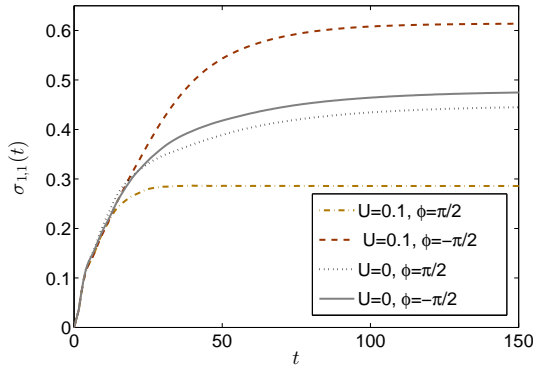}}}
\caption{Occupation of dot '1' in a system with a vertical mirror symmetry using
the parameters of Fig. \ref{FE1},
$N_s=3$, $\delta t=0.8$.
} \label{FE5}
\end{figure}

%==================================================

%===============
\subsection{Model II}

\begin{figure}[t]
{\hbox{\epsfxsize=80mm \epsffile{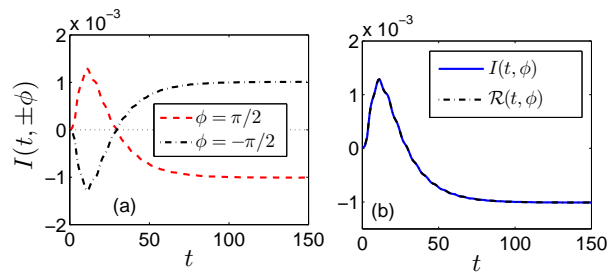}}}
\caption{Model II with an out-of-equilibrium FE.
(a)
A finite drag current is induced in the interferometer (set in equilibrium $\Delta \mu=0$),
a result of its  coupling to a nonequilibrium FE with $\Delta\mu_{F}=1$.
Panel (b) demonstrates that the current in the interferometer
is missing a linear response contribution (and all other odd powers of voltage).
%Parameters are the same as in Fig. \ref{FE1}, but $\Delta\mu_{F}=1$,
%applied in a symmetric form.
$\epsilon_{1,2}=0$, $\gamma_{\nu,n}=0.1$, $\gamma_{\pm}=0.2$, $\epsilon_p=0.1$, $U=0.05$, and
$\beta=10$ (facilitating convergence), $\delta t=0.8$, $N_s=3$.
} \label{FE4}
\end{figure}

Here we simulate with INFPI the dynamics of model II, an AB interferometer coupled to a 
dissipative environment.
%Results are compared to predictions from the voltage probe phenomenological technique \cite{ABs}.
The FE provides a mechanism for elastic and inelastic scattering of electrons on dot '1', and it
dissipates energy from the AB unit. We focus on an equilibrium environment, $\mu_{+}=\mu_-$;
in Fig. \ref{FE4} we further address the role of a nonequilibrium FE on transport symmetries
within the AB interferometer. It should be noted that 
with dot '1' coupled to the FE,
the model lacks a horizontal 
mirror symmetry by construction. 
%As we show
%below, magnetosymmetries in the form (\ref{eq:Mfs}) are obeyed if the model maintains its vertical mirror
%symmetry.

%Observations from numerically exact simulations are compared to a phenomenological treatment,
%the probe technique.
%
%In our simulations the dots are coupled identically to the $L$ or $R$ metals, thus we use
% the short notation $\gamma_{\nu}\equiv\gamma_{\nu,1}=\gamma_{\nu,2}$.

Model II with an equilibrium FE may be mimicked by a noninteracting model
with dot '1' connected to a voltage probe. 
The voltage probe is a metal terminal;  its 
parameters are set so as the net charge current directed towards it, from the AB systems, vanishes. It
provides a mechanism for implementing dissipative inelastic scattering of electrons in the interferometer,
while allowing one to work in the Landauer formalism of noninteracting electrons.
In Ref. \cite{ABs} we used this machinery 
and proved that in systems with a
vertical mirror symmetry even and odd conductance components of the
 charge current (\ref{eq:R})-(\ref{eq:D}) obey in steady-state the relations (\ref{eq:Mfs}).
When spatial asymmetry in the form $\gamma_L=\gamma_{L,1}=\gamma_{L,2}\neq
\gamma_R=\gamma_{R,1}=\gamma_{R,2}$ is introduced, the relations (\ref{eq:Mfs}) are violated \cite{ABs}.
Here we complement the probe analysis  and explore model II with genuine many-body effects:
We confirm the relations (\ref{eq:Mfs}) under a vertical mirror symmetry, study violations of
this equation 
in general cases, and point out that under certain conditions the diode effect, 
missing in the phenomenological probe treatment, shows up in exact simulations.

%The dissipative model II, with an equilibrium environment, has been examined in Ref. \cite{ABs},
%showing that in spatially symmetric systems $\gamma_{\nu,n}=\gamma$ the magnetic field symmetries
%(\ref{eq:MFs}) are obeyed.

In Fig. \ref{FE1} we display the charge current in the interferometer once isolated
($U=0$) or coupled to a FE ($U=0.1$). We confirm that the former, a coherent system, obeys phase
rigidity $I(t,\phi)=I(t,-\phi)$. When the AB setup is coupled to the FE, the transient current
and the steady-state value do not transparently expose any symmetry,
but panels \ref{FE1}(b)-(c) demonstrate that in a geometrically
symmetric setup, $\gamma_{L}=\gamma_{R}$, the symmetries (\ref{eq:Mfs}) are satisfied in  the transient
regime and in the steady-state limit. 
%Deviations from these relations are small, $(\mathcal R(\phi)+\mathcal
%R(-\phi))/2\mathcal R(\phi)\sim (1+i)\times 10^{-6}$.
%%
%Since the real and imaginary parts (the latter reflects computing errors)
%are of the same order, we conclude that deviations from the symmetry (\ref{eq:Mfs}) originates from
%computing errors.

It is important to recall that 
the occupations of the dots in the AB interferometer do not satisfy analogous symmetries, even in 
the isolated $U=0$ limit.
In Fig. \ref{FE5} we display the occupation of dot '1' and show
that, in agreement with analytical results \cite{ABsG},
 $\sigma_{1,1}(t,\phi)$ does not satisfy a phase symmetry away from the symmetric point.
%When dot '1' is further coupled to the FE this  is enhanced
%since the bare energy $\epsilon_1$ further shifts away from $\epsilon_2$.  
Note that at short time,
$\gamma_{\nu}t\lesssim 0.5$, the current and the occupation of dot '1' are insensitive to both interactions
and the magnetic phase. When $U$ is turned on, the case with $\phi=\pi/2$ approaches steady-state significantly
faster than the  opposite $\phi=-\pi/2$ situation. This is reflected in both the occupation dynamics and the current.

\begin{figure}[htpb]
{\hbox{\epsfxsize=80mm \epsffile{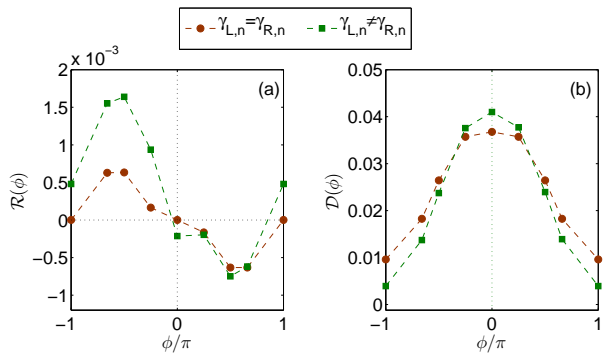}}}
\caption{Steady-state values of $\mathcal R$ and $\mathcal D$ in Model II,
($\circ$) symmetric $\gamma_{\nu,n}=0.2$, and ($\square$) asymmetric
$\gamma_{L,n}=0.05\neq\gamma_{R,n}=0.2$ setups.
Other parameters are the same as in Fig. \ref{FE1}.
}
\label{FE2}
\end{figure}

\begin{figure}[htbp]
{\hbox{\epsfxsize=80mm \epsffile{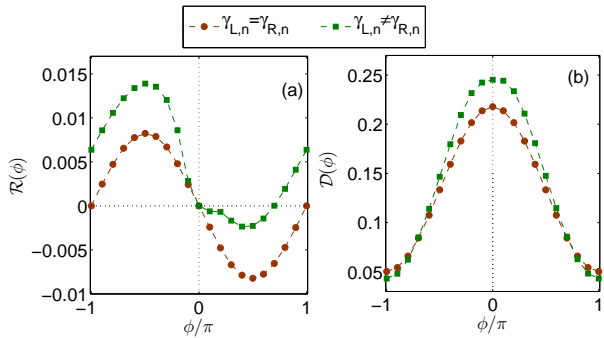}}}
\caption{Model II, with the FE replaced by a voltage probe coupled to dot '1'.
The probe equations are detailed in Ref. \cite{ABs};
($\circ$) symmetric $\gamma_{\nu,n}=0.2$, and ($\square$) asymmetric
$\gamma_{L,n}=0.05\neq\gamma_{R,n}=0.2$ setups.
Parameters are the same as in Figs. \ref{FE1}-\ref{FE2}, besides the temperature which
is set at $1/\beta=25$ and
$\gamma_p=0.05$.
} \label{FE3}
\end{figure}

In Fig. \ref{FE2} we present steady-state data for $\mathcal R$ and $\mathcal D$. We demonstrate the
validity of Eq. (\ref{eq:Mfs}) in junctions with a vertical mirror symmetry, and its violation in
general situations. Note that $\mathcal D(\phi)\neq \mathcal D(-\phi)$ under a spatial asymmetry, 
but deviations are small, for example $\mathcal D(\pi/2)-\mathcal D(-\pi/2)\sim 1.5 \times 10^{-4}$.
We compare these results to 
the probe technique as described in Ref. \cite{ABs}. The coupling of dot
'1' to the probe (hybridization strength $\gamma_p$) does not directly correspond to the capacitive coupling
$U$, thus we can only make a qualitative comparison here. Results are
displayed in Fig. \ref{FE3}. Note that we
used here a higher electronic temperature, $1/\beta=25$, to facilitate convergence. It
was shown in Ref. \cite{ABs} that an increase of the metals' temperature only leads to a weaker
visibility of the current with the magnetic flux, but it does not alter the oscillation of the 
current with flux.
%new qualitative features were not introduced.

%Figs. \ref{FE2} and \ref{FE3} expose
%the role of inelastic scattering processes of electrons in the AB interferometer:
%Fig. \ref{FE2} displays results with genuine many-body effects, Fig. \ref{FE3} is based on a phenomenological treatment.
Comparing  Fig. \ref{FE2} (INFPI) to Fig. \ref{FE3} (probe), we observe that
the probe technique provides qualitative correct features.
However, in asymmetric setups it brings  $\mathcal R(\phi=0)=0$ \cite{ABs}. 
In contrast,
INFPI yields a nonzero value for $\mathcal R(\phi=0)$, see Fig. \ref{FE2}. 
This disagreement has an important implication:
The phenomenological probe approach predicts that in the absence of a magnetic flux the
AB system {\it cannot} act as a diode, though an asymmetry is introduced and (effective) many-body interactions
are playing a role. In contrast, INFPI simulations show that the system can act as a diode at zero
flux if $\gamma_{L}\neq \gamma_{R}$.

We now explore the role of nonequilibrium effects in the FE, $\Delta\mu_{F}=\mu_+-\mu_-\neq 0$.
As long as we keep the interferometer biased $\Delta \mu\neq 0$ we recover the symmetries
as before, reaching the dynamics as in Fig. \ref{FE1}.
Transport symmetries are thus unaffected by the nonequilibrium environment, and this could be justified by
noting that in our model dot '1' is coupled to a number operator in the FE, as in Ref.
\cite{Saito08}, rather than to scattering states \cite{QPC2}.

%=================================================

%=======================

%==================================================

Model II further allows us to explore the development of the ``Coulomb drag current" in the 
interferometer, a result of its coupling to the FE.  
This effect has important implications in nanoscale electronic junctions:
When placing two quantum wires (with independent contacts) into a close proximity,
a ``drive current" passing through one conductor can induce
a ``drag current" in the other wire, a result of Coulomb (capacitive) interactions between 
charges in the two wires,
see e.g. Ref. \cite{Reno} for an experimental demonstration. 
In Fig. \ref{FE4} we explore this effect using INFPI:
The interferometer is unbiased, $\Delta \mu=0$, but we voltage bias the FE.
We show that in a system with a vertical mirror symmetry %($\gamma_{\nu,n}=0.1$ and $\gamma_{\pm}=0.2$),
the drag current is nonzero, an odd function of the magnetic flux. 
We can drive a positive or a negative current in the AB interferometer;
% $\Delta \mu_F$
the directionality is induced here through the magnetic flux,
not the hybridization coefficients as in other works \cite{DragB10}. 
Furthermore, by plotting in panel (b) the measure 
$\mathcal R=[I(t,\phi,\Delta\mu_F)+I(t,\phi,-\Delta \mu_F)]/2$
we confirm that the current includes only even powers in
$\Delta\mu_{F}$, missing altogether a linear response term.
%$I(t,\phi)=\sum_{k=1,2,...}G_{2k}(t,\phi)\Delta\mu_F^{2k}$.

We emphasize that the drag current observed here does not emerge from the transfer of momentum 
between charges, rather, we harness here charge fluctuations in the FE. 
It is thus expected that an unbiased-thermal FE
could induce a net current in a centrosymmetric AB interferometer, if the magnetic flux is nonzero
\cite{DragO}. However,  
this situation cannot be explored at present by the INFPI technique since its 
 convergence requires a large voltage biasing or high temperatures, with the memory 
time approximately given by $\tau_c=\delta t N_s \sim min\{1/\Delta \mu_{F}, \beta\}$. 

The Coulomb drag effect has been examined so far by breaking the spatial symmetry
using uneven contacts, adopting phenomenological rate equations
or perturbative treatments, see e.g. Refs. \cite{DragB07,DragT,DragB10}.
Our work here is a first step towards the exploration of this many-body phenomenon with 
 a broken time reversal symmetry, by means of an exact numerical tool.

We summarize our main observations for model II:
(i) The relations (\ref{eq:Mfs}) are satisfied
in the transient regime and in the steady-state limit when the vertical mirror symmetry is obeyed.
The approach to steady-state depends on the magnetic flux.
(ii) The probe technique, an effective mean,
provides the correct features for $\mathcal R$ and $\mathcal D$, but it predicts no dc rectification
current in the absence of a magnetic flux, for $\gamma_L\neq \gamma_R$.
(iii) The FE may generate a positive or a negative drag current in an unbiased 
centrosymmetric AB interferometer, given a nonzero magnetic phase in the system.
%Whether the FE is set in equilibrium or away from equilibrium does not affect the observed symmetries.

%=======================

%===========================

\section{Summary}

We examined the double-dot AB interferometer with controlled many-body effects, either
internal, between electrons on the dots (Model I), or between the AB electrons 
and a dissipative environment (Model II). 
Using a flexible numerically exact tool, we studied the transient and the steady-state
characteristics of the charge current in the AB system. We validated magnetic field symmetries of nonlinear
transport when the system preserves horizontal or vertical mirror symmetries. Transport asymmetries were
displayed and quantified in general geometries. 
Applications beyond the mean field level were exemplified, including
a charge diode, charge sensing, and the Coulomb drag current.

Earlier studies of Magnetotransport properties were limited to steady-state situations, 
mostly analyzed at the mean-field level. 
Here, we studied a double-dot AB interferometer with genuine many-body interactions, and we
simulated it with a numerically exact tool. The comparison to effective treatments, Hartree MF and the probe
technique, reveals that these simplified methods capture correctly transport symmetries, though 
magnitudes of nonlinear terms may substantially deviate from the exact limit.

%Our simulations were preformed with a deterministic numerical techniques,
%were errors are identically accumulated for $\pm\phi$ simulations, benefitical for
%testing symmetry relations.
%It is of interest to extend our analysis and connect an AB interferometer directly
%to a nonequilibrium object, for example, a quantum point contact,
%and further study symmetries in the nonlinear regime.

In future studies we plan to examine other many-body models, for example, a nanojunction coupled to internal
vibrations. This will be done using INFPI \cite{IFs} and other 
perturbative-analytical and numerical schemes such as the Green's function technique \cite{Hod,Misha} 
and Quantum master equations \cite{QME,Hartle}. 
Such an analysis would not only resolve transport behavior,
but further serve as a critical test for examining the consistency of analytical and numerical 
truncation schemes \cite{Eran}. 
Other ideas involve a detailed analysis of the Coulomb drag effect, harnessing (hot) thermal charge 
fluctuations to drive a dc current in the interferometer \cite{DragB11}. Finally, we plan to
study  symmetries of the thermoelectric current using both the phenomenological probe 
technique and INFPI, with
the objective to suggest means for increasing heat to work conversion efficiency in nonlinear situations
\cite{Sanchez13PRL, Sanchez13, Whitney}.

%===================================================================
\acknowledgments 

The work of SB has been supported by the Early Research Award of DS,
the Martin Moskovits Graduate Scholarship in Science and Technology,
 and the Lachlan Gilchrist Fellowship Fund.
DS acknowledges support from the Natural Science and Engineering Research
Council of Canada.

%=====================================================================

%==========================

%-----------------------------


\begin{thebibliography}{32}
\bibitem{comment2}
The data for $\delta t=0.6$ and $\delta t=1$ reach a different plateau at
large $\tau_c$, see Fig. \ref{ModelIC} (c).
Given the accumulation of the Trotter error, we take below the values
obtained with the smallest time step, $\delta t=0.6$, 
to represent INFPI's numerical soltuion.
More extensive simulations, with $\delta t\sim 0.5$ and $N_s=2-10$, 
are necessary to achieve a more accurate value for $\mathcal R$.
\bibitem{comment}
INFPI simulations allow us to explore the quasi steady-state regime, limited by the recurrence time which is
 dictated by the finite discretization of the fermionic baths.
Numerical simulations in Fig. \ref{ModelIssI} are limited to certain phases since 
the relaxation time towards steady-state significantly 
grows when the magnetic phase is small, $\phi/\pi<1/2$.
Indeed, in Ref. \cite{Salil11} we had shown analytically
that the two-dots dynamics is controlled by rate constants of the 
form $[1\pm\cos(\phi/2)]$;  the rate proportional to $[1-\cos(\phi/2)]$ apparently 
controls the dynamics of $\mathcal R(t,\phi)$.
\end{thebibliography}
\end{document}